\documentclass[submission,copyright,creativecommons]{eptcs}
\usepackage{tikz,wasysym}
\usepackage{ntheorem}
\usetikzlibrary{automata,positioning}
\usetikzlibrary{automata, arrows.meta, positioning}
\usepackage{pgf}
\usetikzlibrary{positioning,automata}
\theoremseparator{.}

\usepackage{tikz}
\usepackage{subcaption}
\usepackage{amsfonts}
\usepackage{amssymb}
\usepackage{verbatim}

\theoremstyle{plain}
	\newtheorem{theorem}{Theorem}
	\newtheorem{notn}[theorem]{Notation}
	\newtheorem{lemma}[theorem]{Lemma}
	\newtheorem{proposition}[theorem]{Proposition}
	\newtheorem{corollary}[theorem]{Corollary}

\theoremstyle{definition}

	\newtheorem{remark}[theorem]{Remark}

\providecommand{\event}{AFL 2023} 
\providecommand{\urlalt}[2]{\href{#1}{#2}}
\providecommand{\doi}[1]{doi:\urlalt{http://dx.doi.org/#1}{#1}}

\usepackage{epsfig,float}
\usepackage{pdfsync}
\usepackage{hyperref}
\usepackage{epstopdf}
\usepackage{graphicx}
\usepackage{epic}
\usepackage{eepic}
\usepackage{bm}

\renewcommand{\le}{\leqslant}

\newcommand{\ol}{\overline}
\newcommand{\eps}{\varepsilon}

\newcommand{\Sig}{\Sigma}

\tikzset{white border/.style={preaction={draw,white,line width=4pt}}}

\usepackage{iftex}

\ifpdf
  \usepackage{underscore}         
  \usepackage[T1]{fontenc}        
\else
  \usepackage{breakurl}           
\fi

\title{Duality of Lattices Associated to Left and Right Quotients}
\author{Jason Bell\thanks{Supported by NSERC grant RGPIN RGPIN-2022-02951.}
\institute{Department of Pure Mathematics\\
University of Waterloo, Canada}
\email{jpbell@uwaterloo.ca}
\and
Daniel Smertnig\thanks{Supported by the Austrian Science Fund (FWF): P 36742-N.}
\institute{Institute for Mathematics and Scientific Computing \\ University of Graz, Austria}
\email{daniel.smertnig@uni-graz.at}
\and
Hellis Tamm\thanks{Supported by the Estonian Research Council grant PRG1210.}
\institute{Department of Software Science \\
  Tallinn University of Technology, Estonia}
\email{hellis@cs.ioc.ee}
}
\def\titlerunning{Duality of Lattices}
\def\authorrunning{J. Bell, D. Smertnig, \& H. Tamm}
\begin{document}
\maketitle

\begin{abstract}
We associate lattices to the sets of unions and intersections of left and right quotients of a regular language.
For both unions and intersections, we show that the lattices we produce using left and right quotients are dual
to each other. We also give necessary and sufficient conditions for these lattices to have maximal possible complexity.
\end{abstract}

\tikzset{white border/.style={preaction={draw,white,line width=4pt}}}

\section{Introduction}
\label{sec:intro}

Within the study of formal languages, a common theme is associating invariants that provide a measure of complexity of the language.  A key example of this type is the entropy of languages (cf. Chomsky and Miller \cite{CM}), which gives a measure of their growth.  

When one restricts to regular languages, one of the most essential notions of complexity comes from the observation that, given a finite alphabet $\Sigma$, a language $L\subseteq \Sigma^*$ is regular if and only if the number of its distinct left quotients is finite, where the left quotient of $L$ by a word $w\in \Sigma^*$ is the language
$$w^{-1}L=\{x\in\Sig^*\colon wx\in L\}.$$  In this sense, the number of distinct left quotients of a regular language provides a measure of its complexity (see 
the survey article \cite{Br2} and references therein for more on quotient complexity).
One can analogously define the right quotient of a language $L$ by a word $v\in\Sig^*$ to be the language 
$$Lv^{-1}=\{u\in\Sig^*\colon uv\in L\},$$ and again, $L$ is regular exactly when it has a finite number of distinct right quotients.  In particular, this gives an analogous notion of complexity.  It should be noted, however, that these two notions of complexity do not coincide.  For example, if $\Sigma=\{a,b\}$ and $L = \{\epsilon, a, a^2, ba\}$, then the left quotients of $L$ are the languages $L, \{\epsilon,a\}, \{a\}, \{\epsilon\}, \varnothing$, while the right quotients are the languages $L, \{\epsilon,a,b\}, \{\epsilon\}, \varnothing$.  

The purpose of this paper is to show that when one instead forms lattices\footnote{A lattice is simply a partially ordered set $(\Lambda,\le)$ with the property that finite subsets have unique least upper bounds and unique greatest lower bounds; thus lattices have a join, $\vee$, and meet, $\wedge$, which are binary operations corresponding to taking respectively the least upper bound and greatest lower bound of two elements of $\Lambda$.} associated with the left and right quotients of a regular language in a natural way, then a duality arises that provides a left-right symmetric measure of the complexity of the language in terms of quotients.  To make this precise, we observe that if $L\subseteq \Sigma^*$ is a regular language with left quotients $L_0,\ldots , L_{n-1}$ and right quotients $R_0, \ldots ,R_{m-1}$, then one can consider the following four lattices.

\begin{itemize}
\item[$\bullet$] The \emph{left quotient union lattice}, ${\rm Latt}(L, \cup, \mathsf{L})$:
\vskip 2mm
the lattice whose elements are all sets that can be formed by taking a (possibly empty) union of left quotients $L_0,\ldots , L_{n-1}$.
\vskip 2mm
\item[$\bullet$]The \emph{right quotient union lattice}, ${\rm Latt}(L, \cup, \mathsf{R})$:
\vskip 2mm
the lattice whose elements are all sets that can be formed by taking a (possibly empty) union of right quotients $R_0,\ldots , R_{m-1}$.
\vskip 2mm
\item[$\bullet$] The \emph{left quotient intersection lattice}, ${\rm Latt}(L, \cap, \mathsf{L})$: 
\vskip 2mm
the lattice whose elements are all sets that can be formed by taking a (possibly empty) intersection of left quotients $L_0,\ldots , L_{n-1}$.
\vskip 2mm
\item[$\bullet$] The \emph{right quotient intersection lattice},  ${\rm Latt}(L, \cap, \mathsf{R})$:
\vskip 2mm
the lattice whose elements are all sets that can be formed by taking a (possibly empty) intersection of right quotients $R_0,\ldots , R_{m-1}$.
\end{itemize}
We observe that the above sets are partially ordered by inclusion and have a join operation, $\vee$, and a meet operation, $\wedge$.  In the case of ${\rm Latt}(L, \cup, \mathsf{L})$ and ${\rm Latt}(L, \cup, \mathsf{R})$, the join of $A$ and $B$ is the union and the meet is the union of all elements of the set that are contained in $A\cap B$, where an empty union is the empty set.  These two lattices have a unique smallest element (the empty set, which is the empty union) and a unique largest element, consisting of the union of all left (respectively right) quotients.

Similarly, in the case of ${\rm Latt}(L, \cap, \mathsf{L})$ and ${\rm Latt}(L, \cap, \mathsf{R})$, the meet is just the intersection and the join of two intersections of quotients, $A$ and $B$, is the intersection of all quotients that contain the union $A\cup B$, where an empty intersection is taken to be $\Sigma^*$.  Then these two lattices again have a unique maximal element $\Sigma^*$ and a unique minimal element given by the intersection of all left (respectively right) quotients.  

As a simple example, consider again the finite regular language $L=\{\epsilon,a,a^2,ba\} \subseteq \{a,b\}^*$.  Then the left quotients are the languages
\begin{equation}L_0=\{\epsilon,a,a^2,ba\},\ L_1=\{a\},\ L_2=\{\epsilon, a\},\ L_3=\{\epsilon\},\ L_4=\varnothing
\label{eq:leftquotient}
\end{equation} while the right 
quotients 
are 
\begin{equation}
\label{eq:rightquotient}
R_0=\varnothing,\ R_1= \{\epsilon\},\
R_2=\{\epsilon, a,b\},\  R_3=\{\epsilon,a,a^2,ba\}
\end{equation}
and we construct the four lattices we consider in this paper from these left and right quotients of $\{\epsilon,a,a^2,ba\}$ in Figures \ref{fig:L2} and \ref{fig:L1}.

\begin{figure}[hbt]
\begin{center}

\begin{tikzpicture}[x=1.4cm,y=1.8cm]
\node at (0,0)    (I)  {$\varnothing$};
\node at (-1,.6)   (A1) {$\{\epsilon\}$};
\node at (1,.6)   (A2) {$\{a\}$};
\node at (0,1.2) (B) {$\{\epsilon ,a\}$};
\node at (0,1.8) (C1) {$\{\epsilon, a, a^2, ba\}$};

\node at (5,1.8)    (S)  {$\{\epsilon,a,a^2,b,ba\}$};
\node at (4,1.2)   (T1) {$\{\epsilon,a,b\}$};
\node at (6,1.2)   (U1) {$\{\epsilon, a, a^2,ba\}$};
\node at (5,.6) (U2) {$\{\epsilon\}$};
\node at (5,0)    (W)  {$\varnothing$};
\draw (S) -- (T1);
\draw (S) -- (U1);
\draw (T1) -- (U2);
\draw (U1) -- (U2);
\draw (U2) -- (W);
\draw (I) -- (A1);
\draw (I) -- (A2);
\draw (A1) -- (B);
\draw (A2) -- (B);
\draw (B) -- (C1);
\end{tikzpicture}

\end{center}
\caption{The lattices ${\rm Latt}(L, \cup, \mathsf{L})$ (left) and  ${\rm Latt}(L, \cup, \mathsf{R})$ (right) for $L=\{\epsilon,a,a^2,ba\}$.} 
\label{fig:L2}
\end{figure}

\begin{figure}[hbt]
\begin{center}
\begin{tikzpicture}[x=1.4cm,y=1.8cm]
\node at (0,0)    (I)  {$\varnothing$};
\node at (-1,.6)   (B1) {$\{\epsilon\}$};
\node at (1,.6) (B2) {$\{a\}$};
\draw (I) -- (B1);
\draw (I) -- (B2);
\node at (0,1.2)    (C2) {$\{\epsilon, a\}$};
\draw (B1) -- (C2);
\draw (B2) -- (C2);
\node at (0,1.8)  (C3) {$\{\epsilon,a,a^2,ba\}$};
\draw (C2) -- (C3);
\node at (0,2.4)    (G)  {$\Sigma^*$};
\draw (C3) -- (G);
\node at (5,2.4)    (S)  {$\Sigma^*$};
\node at (4,1.8)   (T1) {$\{\epsilon, a,b \}$};
\node at (6,1.8) (T2) {$\{\epsilon,a,a^2,ba\}$};
\draw (S) -- (T1);
\draw (S) -- (T2);
\node at (5,1.2)   (U1) {$\{\epsilon,a\}$};
\draw (T1) -- (U1);
\draw (T2) -- (U1);
\node at (5,.6)    (V2) {$\{\epsilon\}$};
\node at (5,0)    (W)  {$\varnothing$};
\draw (W) -- (V2);
\draw (V2) -- (U1);
\end{tikzpicture}
\end{center}
\caption{The lattices ${\rm Latt}(L, \cap, \mathsf{L})$ (left) and  ${\rm Latt}(L, \cap, \mathsf{R})$ (right) for $L=\{\epsilon,a,a^2,ba\}$.} 
\label{fig:L1}
\end{figure}

Figures \ref{fig:L2} and \ref{fig:L1} hint at an unexpected duality. We recall that if $\Lambda$ is a lattice, then we have a dual lattice  $\Lambda^{*}$, which is $\Lambda$ as a set, but where the partial order on $\Lambda$ is reversed and the meet and join are exchanged.  Intuitively, one can think of this as simply taking the lattice $\Lambda$ and writing it ``upside-down''; in particular, the two lattices in Figure \ref{fig:L2} are duals of each other and similarly for the two lattices in Figure \ref{fig:L1}.  

We recall that two lattices $\Lambda$ and $\Lambda'$ are isomorphic (written $\Lambda\cong \Lambda'$) if there is a bijection $f:\Lambda \to \Lambda'$ such that $x<y$ in $\Lambda$ if and only if $f(x)<f(y)$ in $\Lambda'$ and such that $f(x\vee y) = f(x)\vee f(y)$ and $f(x\wedge y)=f(x)\wedge f(y)$ for all $x,y\in \Lambda$.   
Our main theorem shows that the duality occurring in Figures \ref{fig:L2} and \ref{fig:L1} is part of a general phenomenon. 
\begin{theorem}
Let $L\subseteq \Sigma^*$ be a regular language.  Then we have:
\begin{enumerate}
\item[(a)] ${\rm Latt}(L, \cup, \mathsf{L})$ is isomorphic to the dual lattice of ${\rm Latt}(L, \cup, \mathsf{R})$;
\item[(b)] ${\rm Latt}(L, \cap, \mathsf{L})$ is isomorphic to the dual lattice of ${\rm Latt}(L, \cap, \mathsf{R})$.

\end{enumerate}
 \label{thm:main}
 \end{theorem}
 We note that the isomorphism given in Theorem \ref{thm:main}(b), while not stated, can be obtained from the work of Im and Khovanov \cite{SeongIm}, if one carefully analyzes their constructions.  In particular, it would also be interesting to know whether the isomorphism in Theorem \ref{thm:main}(a) has any relevance to one-dimensional topological theories. 
 
 The outline of this paper is as follows.  In \S\ref{sec:pre} we present basic concepts needed from the theory of finite-state automata.  In \S\ref{sec:atom} we provide an overview of the theory of atoms of regular languages and in \S\ref{sec:rel} we describe a key relationship between quotients and atoms.  In \S\ref{sec:proofa} and \S\ref{sec:proofb} we give the proof of Theorem \ref{thm:main}(a) and (b) respectively.  In \S\ref{sec:semi} we relate our results to Boolean semimodules and describe the duality algebraically. In \S\ref{sec:complexity} we present a brief analysis of when the lattices we construct are of maximal possible complexity, and \S\ref{sec:concl} concludes the paper.

\section{Automata and languages}
\label{sec:pre}

A~\emph{nondeterministic finite automaton (NFA)} is a quintuple 
$$\mathcal{N}=(Q, \Sig, \delta, I,F),$$ where 
$Q$ is a finite, non-empty set of \emph{states}, 
$\Sig$ is a finite non-empty \emph{alphabet}, 
$\delta:Q\times \Sig\to 2^Q$ is the  \emph{transition function},
$I\subseteq  Q$ is the set of  \emph{initial states},
and $F\subseteq Q$ is the set of \emph{final states}.
We can naturally extend the transition function to functions 
$$\delta':Q\times \Sig^*\to 2^Q \qquad \delta'':2^Q\times \Sig^*\to 2^Q,$$
which corresponds to taking elements of $\Sig^*$ as input for our automata and read them left-to-right to determine whether or not they are accepted; we henceforth use $\delta$ to denote all of these functions.

The \emph{left language} of a state $q$ of $\mathcal{N}$ is
\begin{equation}
\{w\in\Sig^* \colon q\in \delta(I,w)\},
\end{equation}
 and
the \emph{right language} of $q$ is 
\begin{equation}
\{w\in\Sig^* \colon \delta(q,w)\cap F\neq\varnothing\}.
\end{equation}
A state $q$ of $\mathcal{N}$ is \emph{reachable} if its left language is non-empty, and
it is \emph{empty} if its right language is empty.
The \emph{language accepted} by an NFA $\mathcal{N}$ is 
$L(\mathcal{N})=\{w\in\Sig^*\colon \delta(I,w)\cap F\neq \varnothing\}$, and we say that 
two NFAs are \emph{equivalent} if they accept the same language. 
The \emph{reverse} of an NFA $\mathcal{N}=(Q, \Sig, \delta, I,F)$ is the NFA 
$\mathcal{N}^{R}=(Q, \Sig, \delta^{R},F,I)$, where $q\in\delta^{R}(p,a)$ if and
only if $p\in\delta(q,a)$ for $p,q\in Q$ and $a\in\Sigma$.  The reverse of an NFA $\mathcal{N}$ accepts the reverse of the language accepted by $\mathcal{N}$.

A \emph{deterministic finite automaton (DFA)} is a quintuple 
$\mathcal{D}=(Q, \Sig, \delta, q_0,F)$, where
$Q$, $\Sig$, and $F$ are as in an NFA, 
$\delta:Q\times \Sig\to Q$ is the transition function, 
and $q_0$ is the initial state.

We recall that a language $L$ is regular if it is accepted by some DFA (or equivalently by an NFA).
It is well known that the left quotients of the language $L$ are precisely the right languages of
the states of a minimal DFA for $L$.
Any NFA $\mathcal{N}$ can be \emph{determinized} by the well-known subset
construction, yielding a DFA $\mathcal{N}^D$ that has only reachable states.  We note that one can iteratively perform the reverse and determinization procedures; indeed, this plays a key role in the fundamental work of Brzozowski \cite{Brz63}, and the following result is a slightly modified version of his work. 

\begin{proposition}
\label{thm:Brz}
If an NFA $\mathcal{N}$ has no empty states and $\mathcal{N}^R$ is deterministic, 
then $\mathcal{N}^D$ is a minimal DFA.
\end{proposition}

We note that by Proposition~\ref{thm:Brz}, for any NFA $\mathcal{N}$,
the DFA $\mathcal{N}^{RDRD}$ is the minimal DFA equivalent to $\mathcal{N}$;
this result is known as Brzozowski's double-reversal method for DFA minimization.

\section{Atoms of a regular language}
\label{sec:atom}

Let $L$ be a non-empty regular language with left quotients $L_0,\ldots, L_{n-1}$.
Given a subset $S\subseteq \{0,\ldots ,n-1\}$ we can form a \emph{left atomic intersection} 
\begin{equation}\label{eq:leftatomic}
I_S:= \left(\bigcap_{i\in S} L_i \right) \cap \left(\bigcap_{j\in \{0,\ldots ,n-1\}\setminus S} \ol{L_j}\right),\end{equation} 
where $\ol{L_i}$ is the complement of $L_i$ in $\Sig^*$.

A non-empty left atomic intersection is called a \emph{left atom} of $L$ \cite{BrTa14}.\footnote{In the literature, one generally just uses the term \emph{atom} when speaking of what we call left atoms. However, to achieve our duality results it is convenient to use the adjective \emph{left} when speaking of atoms obtained from left quotients.}
 A left atom is \emph{initial} if it is contained in $L$ and it is \emph{final} if it contains the empty word~$\eps$.  
There is exactly one final left atom; namely the atom $I_T$ where $T$ is the set of $i$ for which $\epsilon\in L_i$.

If $\ol{L_0}\cap \cdots \cap \ol{L_{n-1}}$ is a left atom, then it is called
the \emph{negative} atom, with all other left atoms called \emph{positive}.
Thus left atoms of $L$ are pairwise disjoint languages uniquely determined by $L$ and they define a partition of $\Sig^*$.

One can do a similar construction using right quotients:
if $R_0,\ldots ,R_{m-1}$ are the right quotients of $L$, then for each subset $T\subseteq \{0,\ldots ,m-1\}$ we can form a right atomic intersection 
\begin{equation}
J_T:= \left(\bigcap_{i\in T} R_i \right) \cap \left(\bigcap_{j\in \{0,\ldots ,m-1\}\setminus T} \ol{R_j}\right),\end{equation} 
and we define right atoms of $L$ to be the non-empty right atomic intersections.

We note that the left (resp., right) atoms of a language $L$ are precisely the atoms of the Boolean algebra
(regarded as a partially ordered set), generated by the left (resp., right) quotients of $L$.

As an example, if we take $L=\{\epsilon,a,a^2,ba\}$ then the left atoms in this case are given by the partition 
\begin{equation}
\label{eq:leftatom}
A_0=\ol{\{\epsilon,a,a^2,ba\}},\ A_1=\{a^2,ba\},\ A_2=\{a\},\ A_3=\{\epsilon\}
\end{equation} of $\Sig^*$. These left atoms can be expressed as left atomic intersections as follows:
\begin{equation}\label{i1}\ol{\{\epsilon,a,a^2,ba\}} = \ol{\{\epsilon,a,a^2,ba\}} \cap  \ol{ \{a\}} \cap
\ol{\{\epsilon,a\}}\cap \ol{ \{\epsilon\}} \cap  \ol{\varnothing} = \ol{L_0}\cap \ol{L_1}\cap \ol{L_2}\cap \ol{L_3}\cap \ol{L_4},
\end{equation}
\begin{equation} \label{i2}
\{a^2,ba\}=  {\{\epsilon,a,a^2,ba\}} \cap \ol{ \{a\}} \cap
\ol{\{\epsilon,a\}}\cap  \ol{ \{\epsilon\}} \cap  \ol{\varnothing}= {L_0}\cap \ol{L_1}\cap \ol{L_2}\cap \ol{L_3}\cap \ol{L_4},
\end{equation}
\begin{equation}\label{i3}
\{a\}  =  {\{\epsilon,a,a^2,ba\}} \cap { \{a\}} \cap
{\{\epsilon,a\}}\cap  \ol{ \{\epsilon\}} \cap  \ol{\varnothing}= {L_0}\cap {L_1}\cap {L_2}\cap \ol{L_3}\cap \ol{L_4},
\end{equation}
and
\begin{equation}\label{i4}
\{\epsilon\}= {\{\epsilon,a,a^2,ba\}} \cap \ol{ \{a\}} \cap
{\{\epsilon,a\}}\cap  { \{\epsilon\}} \cap  \ol{\varnothing}= {L_0}\cap \ol{L_1}\cap {L_2}\cap {L_3}\cap \ol{L_4}.
\end{equation}

 Here, the left atom $\ol{\{\epsilon,a,a^2,ba\}}$ is negative, while the remaining left atoms are both positive and initial and the left atom $\{\epsilon\}$ is the unique final atom.  
 
 On the other hand, the right atoms are given by the partition \begin{equation}
\label{eq:rightatom}
B_0 = \{ \epsilon \},\ B_1= \{b\},\ B_2=\{a\},\ 
B_3= \{ba,a^2\},\ B_4= \ol{ \{\epsilon,a,a^2, b,ba\}},\end{equation}
and they are obtained as right atomic intersections as
\begin{equation}\label{e1'}
 \{ \epsilon \} = \ol{\varnothing}\cap \{\epsilon\}\cap \{\epsilon,a,b\}\cap \{\epsilon,a,a^2,ba\}= \ol{R_0}\cap {R_1} \cap {R_2} \cap {R_3},
\end{equation}

\begin{equation}\label{e2'}
 \{ b \} =  \ol{\varnothing}\cap \ol{\{\epsilon\}}\cap 
 \{\epsilon,a,b\}\cap \ol{\{\epsilon,a,a^2,ba\}} = \ol{R_0}\cap \ol{R_1} \cap {R_2} \cap \ol{R_3},
\end{equation}

\begin{equation}\label{e3'}
 \{ a \} = \ol{\varnothing}\cap \ol{\{\epsilon\}}\cap 
 \{\epsilon,a,b\}\cap {\{\epsilon,a,a^2,ba\}}  = \ol{R_0}\cap \ol{R_1} \cap {R_2} \cap {R_3},
\end{equation}

\begin{equation}\label{e4'}
 \{ ba, a^2 \} = \ol{\varnothing}\cap \ol{\{\epsilon\}}\cap 
\ol{ \{\epsilon,a,b\}}\cap {\{\epsilon,a,a^2,ba\}} = \ol{R_0}\cap \ol{R_1} \cap \ol{R_2} \cap {R_3},
\end{equation}
and
\begin{equation}\label{e5'}
\ol{ \{\epsilon,a,a^2, b,ba\}} =  \ol{\varnothing}\cap \ol{\{\epsilon\}}\cap 
\ol{ \{\epsilon,a,b\}}\cap \ol{\{\epsilon,a,a^2,ba\}} = \ol{R_0}\cap \ol{R_1} \cap \ol{R_2} \cap \ol{R_3}.
\end{equation}

We note that every left quotient of $L$ (including $L$ itself) is a (possibly empty) union of left atoms and similarly every right quotient is a union of right atoms.

It is well known that left quotients of $L$ are in a one-to-one correspondence with
the equivalence classes of the \emph{Nerode right congruence} $\equiv_{L}$
of $L$~\cite{Ne58} defined as follows: for $x,y\in\Sigma^*$, $x \equiv_{L} y$ 
if for every $v\in\Sigma^*$, $xv\in L$ if and only if $yv\in L$.
Left atoms of $L$ are the classes of the \emph{left congruence}
${}_{L}{\equiv}$ of $L$:
for $x,y\in\Sigma^*$, $x \mathbin{{}_{L}{\equiv}} y$ if for every $u\in\Sigma^*$, 
$ux\in L$ if and only if $uy\in L$~\cite{Iv16}.
Also, right quotients are in a one-to-one correspondence with the equivalence classes of the left congruence.

Let $A_{0},\ldots , A_{{m-1}}$ denote the left atoms of $L$ where we index so that $A_{m-1}$ is the final atom, and let $I$ denote
the set of initial atoms.

The \emph{\'atomaton} $\mathcal{A}$ of $L$ is the NFA whose set of states is the set \begin{equation}
S = \{s_0,\ldots ,s_{m-1}\},
\end{equation} which can be thought of as parameterizing the set of left atoms of $L$.  More precisely,
we take $$\mathcal{A}=(S,\Sig,\alpha,I,\{s_{m-1}\}),$$ where
$s_j \in \alpha(s_i, a)$ if and only if $A_{j} \subseteq a^{-1}A_{i}$,
for $i,j\in\{0,\ldots,m-1\} $ and $a\in\Sig$.  (We refer the reader to \cite{BrTa14} for further details on \'atomata.)

In the running example in which we take $L=\{\epsilon, a,a^2,ba\}$, by Equation (\ref{eq:leftatom}), the left atoms are the sets 
$$A_0=\ol{\{\epsilon,a,a^2,ba\}},\ A_1= \{a^2,ba\},\ A_2= \{a\},\ A_3=\{\epsilon\},$$ and we see that the \'atomaton associated to $L$ is given in Figure \ref{fig:atomaton-DFA} on the left, where the states $s_1,s_2,s_3$ are initial.

\begin{figure}
\begin{center}
\begin{minipage}{.4\textwidth}
\begin{tikzpicture} [node distance = 2.3cm, on grid, auto,initial text=]
\node (q0) [state] {$s_0$};
\node (q1) [state, initial, above = of q0] {$s_1$};
\node (q2) [state, initial above, right = of q1] {$s_2$};
\node (q3) [state, initial, initial where=right, accepting, below = of q2] {$s_3$};
\path [-stealth]
    (q0) edge node {$a,b$}   (q1)
    (q0) edge node {$b$}  (q3)
       (q0) edge [loop below] node {$a,b$}   (q0)
       (q1) edge node {$a,b$} (q2)
       (q2) edge node {$a$} (q3);
\end{tikzpicture}
\end{minipage}
\begin{minipage}{.4\textwidth}     
\begin{tikzpicture} [node distance = 2.3cm, on grid, auto,initial text=]
\node (q0) [state, initial, accepting] {$q_0$};
\node (q1) [state,  above right = of q0] {$q_1$};
\node (q2) [state, accepting, below right = of q0] {$q_2$};
 \node (q3) [state, accepting, right = of q1] {$q_3$};
 \node (q4) [state, right = of q2] {$q_4$};
\path [-stealth]
    (q0) edge node {$b$}   (q1)
  (q0) edge node {$a$}  (q2)
       (q1) edge node {$a$}   (q3)
       (q1) edge node  {$b$} (q4)
           (q2) edge node {$b$} (q4)
             (q2) edge [bend left] node {$a$} (q3)
       (q3) edge node {$a,b$} (q4)
    (q4) edge [loop right]  node {$a,b$}();
\end{tikzpicture}
\end{minipage}
\end{center}
\caption{The \'atomaton (left) and the minimal DFA (right) for $L=\{\epsilon, a,a^2,ba\}$.}
\label{fig:atomaton-DFA}
\end{figure}

Observe that if we adopt the labelling given in Equations (\ref{eq:rightquotient}) and (\ref{eq:leftatom}), then the right languages of the \'atomaton in Figure 3 are $$A_0=\{a,b\}^*\cdot \left( \{b\}\cup \{a,b\}^2\cdot \{a\}\right)=\ol{\{\epsilon,a,a^2,ba\}}$$ (for the state $s_0$), $A_1=\{a^2,ba\}$ (for the state $s_1$), $A_2=\{a\}$ (for the state $s_2$), and $A_3
=\{\epsilon\}$ (for the state $s_3$), which are precisely the left atoms of the language given in Equation (\ref{eq:leftatom}).
 
On the other hand, the left languages are $R_0=\varnothing$ (for the state $s_0$), $R_1
=\{\epsilon\}$ (for the state $s_1$), $R_2=\{\epsilon, a,b\}$ (for the state $s_2$), $R_3=\{\epsilon, a, a^2, ba\}$ (for the state $s_3$), and these are precisely the right quotients of $L$, as given in Equation (\ref{eq:rightquotient}).
 
In fact, these observations are part of general phenomena, as shown by Brzozowski and Tamm \cite{BrTa14}, which we record in the following proposition.

\begin{proposition}
\label{prop:atoms-rquot} Let $L$ be a non-empty regular language.
Then the following hold:
\begin{enumerate}
\item[(i)] the left quotients of $L$ are precisely the right languages of the minimal DFA accepting $L$;
\item[(ii)] the left atoms of $L$ are precisely the right languages of the \'atomaton $\mathcal{A}$ associated to $L$;
\item[(iii)] the right quotients of $L$ are precisely the left languages of $\mathcal{A}$;
\item[(iv)] the right atoms of $L$ are precisely the left languages of the minimal DFA accepting $L$.

\end{enumerate}
 In particular, we have set bijections
 $$\{\textrm{\it left atoms of }L\} \leftrightarrow \{\textrm{\it right quotients of }L\}$$ and 
  $$\{\textrm{\it right atoms of }L\} \leftrightarrow \{ \textrm{\it left quotients of }L\},$$
where in the first case we view a left atom of $L$ as the right language of a state of $\mathcal{A}$ and then send it to the left language of this state and in the second case we view a right atom of $L$ as the left language of a state of the minimal DFA of $L$ and then send it to the right language of this state.
\end{proposition}
\vskip 2mm
\emph{Proof.} Item ($i$) is well known.  It was shown in~\cite{BrTa14} that the left atoms of a regular language $L$ are precisely the right languages
of the states of the associated \'atomaton, so ($ii$) holds.

A modification of the isomorphism result from \cite{BrTa14} shows that if $\mathcal{D}$ is the minimal DFA accepting $L$ with state set $Q=\{q_0,q_1,\ldots, q_{n-1}\}$, then the \'atomaton, $\mathcal{A}$, associated to $L$ is isomorphic to $\mathcal{D}^{RDR}$ as NFAs, 
via an isomorphism induced by the map which sends a state $s_i\in S$ from the state set of $\mathcal{A}$ to the set
$\{q_j \colon j\in S\}$, where $S\subseteq \{0,\ldots,n-1\}$ has the property that $A_i$ is the left atomic intersection $I_S$.  Since by Proposition~\ref{thm:Brz}, the DFA $\mathcal{D}^{RD}$ is the minimal DFA of the reverse
language of $L$, the left languages of $\mathcal{D}^{RDR}\cong \mathcal{A}$ are exactly the right quotients of $L$, which establishes ($iii$).

Finally, \cite{BrTa14} shows that the reverse NFA of the \'atomaton of $L$ is the minimal DFA of the reverse language of $L$, and so ($iv$) now follows, and the bijections are immediate from ($i$)--($iv$). $\square$
\vskip 3mm

We again consider the regular language $L=\{\epsilon,a,a^2,ba\}$ as an example.
Then the automaton in Figure \ref{fig:atomaton-DFA} on the right is the minimal DFA accepting $L$
with the state set $\{q_0,q_1,q_2,q_3,q_4\}$. 

%

Observe that for this DFA, if we adopt the labellings from Equations (\ref{eq:leftquotient}) and (\ref{eq:rightatom}), the left language of $q_0$ is $B_0=\{\epsilon\}$ and the right language is $L_0=L$; the left language of $q_1$ is the right atom $B_1=\{b\}$ and the right language is the left quotient $L_1=\{a\}$; the left language of $q_2$ is $B_2=\{a\}$ and the right language is $L_2=\{\epsilon,a\}$;  the left language of $q_3$ is $B_3=\{ba,a^2\}$ and the right language is $L_3=\{\epsilon\}$; and finally the left language of $q_4$ is $B_4=(\{ab,b^2\} \cup \{a^2,ba\}\{a,b\})\{a,b\}^*=\ol{\{\epsilon,a,b,a^2,ba\}}$ and the right language is $L_4=\varnothing$.  Similarly, the remarks preceding Proposition \ref{prop:atoms-rquot} give the bijection between left atoms and right quotients.  We record these bijections in Figure \ref{fig:T1}, where $\mathcal{A}$ is the \'atomaton and $\mathcal{D}$ is the DFA from Figure \ref{fig:atomaton-DFA}.
 \begin{figure}
\begin{center}
{\renewcommand{\arraystretch}{1.5}
\begin{tabular}{|c|c|c|}
\hline
State of $\mathcal{D}$ & Left quotient of $\{\epsilon,a,a^2,ba\}$ & Right atom of $\{\epsilon,a,a^2,ba\}$  \\
\hline
$q_0$ & $\{\epsilon,a,a^2,ba\}$ & $\{\epsilon\}$ \\
\hline
$q_1$ & $\{a\}$ & $\{b\}$ \\
\hline
$q_2$ & $\{\epsilon,a\}$ & $\{a\}$ \\
\hline
$q_3$ & $\{\epsilon\}$ & $\{ba,a^2\}$ \\
\hline
$q_4$ & $\varnothing$ & $\ol{\{\epsilon,a,b,a^2,ba\}}$ \\
\hline
\end{tabular}}
\vskip 7mm
{\renewcommand{\arraystretch}{1.5}
\begin{tabular}{|c|c|c|}
\hline
State of $\mathcal{A}$ & Right quotient of $\{\epsilon,a,a^2,ba\}$ & Left atom of $\{\epsilon,a,a^2,ba\}$  \\
\hline
$s_0$ & $\varnothing$ & $\ol{\{\epsilon,a,a^2,ba\}}$ \\
\hline
$s_1$ & $\{\epsilon\}$ & $\{a^2,ba\}$ \\
\hline
$s_2$ & $\{\epsilon,a,b\}$ & $\{a\}$ \\
\hline
$s_3$ & $\{\epsilon,a,a^2,ba\}$ & $\{\epsilon\}$ \\
\hline
\end{tabular}}
\end{center}
  \caption{Tables giving the bijections described in Proposition \ref{prop:atoms-rquot} between left quotients and right atoms and between right quotients and left atoms for the language $L=\{\epsilon, a,a^2,ba\}$.}
  \label{fig:T1}
  \end{figure}

\section{Relationships between quotients and atoms}
\label{sec:rel}
In this section, we give key bijections between left quotients and right atoms and similarly for right quotients and left atoms.

We find it convenient to introduce notation that we will use in proving Theorem \ref{thm:main}.  The main aim of this notation is to capture the isomorphisms described in Proposition \ref{prop:atoms-rquot} and we henceforth adopt this notation in all results we prove.
\begin{notn} We introduce the following notation.
\begin{enumerate}
\item[(i)] We let $L$ be a non-empty regular language in $\Sigma^*$ with $\Sigma$ a finite alphabet.
\item[(ii)] We let $\mathcal{A}$ denote the \'atomaton of $L$ and let $\mathcal{D}$ denote the minimal DFA accepting $L$ on states $q_0,\ldots ,q_{n-1}$.
\item[(iii)] We let $L_0,\ldots ,L_{n-1}$ denote the left quotients of $L$.
\item[(iv)] We let $R_0,\ldots ,R_{m-1}$ denote the right quotients of $L$.
\item[(v)] We let $A_0,\ldots, A_{m-1}$ denote the left atoms of $L$, 
where we index so that $A_i$ corresponds to $R_i$ under the bijection given in Proposition \ref{prop:atoms-rquot}.
\item[(vi)] We let $B_0, \ldots, B_{n-1}$ 
 be the right atoms of $L$, where we index so that $B_i$ corresponds to $L_i$ under the bijection given in Proposition \ref{prop:atoms-rquot}.
 \end{enumerate}
 \label{notn:notn}
\end{notn}

\begin{remark} We note that in our running example where $L=\{\epsilon, a,a^2,ba\}$, this notation is consistent with the labellings given in Equations (\ref{eq:leftquotient}), (\ref{eq:rightquotient}), (\ref{eq:leftatom}), and (\ref{eq:rightatom}), as shown by Figure \ref{fig:T1}.
\end{remark}

The following proposition gives a precise relationship between the left and right quotients
of a regular language $L$ and the left and right atoms of $L$.

\begin{proposition} Let
\label{prop:r-quot}
$i\in \{0,\ldots ,m-1\}$, and let $j\in \{0,\ldots ,n-1\}$. Then
$$R_i=\bigcup_{\{k\colon A_i\subseteq L_k\}} B_k \qquad \textrm{\it and} \qquad
L_j=\bigcup_{\{\ell \colon B_j\subseteq R_{\ell}\}} A_{\ell} .$$
In particular, $A_i\subseteq L_j$ if and only if $B_j\subseteq R_i$.
\end{proposition}
\emph{Proof.}
As noted in the proof of Proposition \ref{prop:atoms-rquot}, the modified argument of \cite{BrTa14} shows that the NFAs $\mathcal{A}$ and $\mathcal{D}^{RDR}$
are isomorphic, with a state $s_i$ in $\mathcal{A}$ corresponding to a set $\{q_i\colon i\in S\}$ for some set $S\subseteq \{0,\ldots ,n-1\}$ with the property that the left atom $A_i$ is the (left) atomic intersection $I_S$ described in Equation (\ref{eq:leftatomic}).  By Proposition \ref{prop:atoms-rquot}, the right quotients of $L$ are the left languages of $\mathcal{D}^{RDR}$ and the right atoms of $L$ are the left languages of $\mathcal{D}$.
Hence, it is clear that the first equality holds.  

The second equality is proved analogously, now using that the left quotients of $L$ are the right languages of $\mathcal{D}$ and the left atoms of $L$ are the right languages of $\mathcal{A}\cong \mathcal{D}^{RDR}$ by Proposition \ref{prop:atoms-rquot}. The ``in particular'' clause follows immediately from these equalities. 
$\square$


\begin{lemma}
\label{prop:U-V} Let $X$ be a union of left atoms of $L$, and let $Y$ be a union of right atoms of $L$. Then we have:
\begin{enumerate}
\item[(1)] $\bigcup_{\{i\colon A_i\not\subseteq X\}}R_i =\bigcup_{\{ j\colon L_j\not\subseteq X\}}B_j,$
\item[(2)] $\bigcap_{\{j\colon A_j\subseteq X\}}R_j=\bigcup_{\{i\colon X\subseteq L_i\}}B_i,$

\item[(3)] $\bigcup_{\{i\colon B_i\not\subseteq Y\}}L_i =\bigcup_{\{ j\colon R_j\not\subseteq Y\}}A_j,$
\item[(4)] $\bigcap_{\{j\colon B_j\subseteq Y\}}L_j=\bigcup_{\{i\colon Y\subseteq R_i\}}A_i.$
\end{enumerate}
\end{lemma}
\emph{Proof.}
We consider the union of the right quotients $U=\bigcup_{A_i\not\subseteq X}R_i$,
corresponding to the left atoms not contained in $X$.

Consider a left quotient $L_j$ that is not contained in $X$.
Then there is some left atom $A_i$ such that $A_i\subseteq L_j$ and
$A_i\not\subseteq X$.
By Proposition \ref{prop:r-quot}, $A_i\subseteq L_j$ gives that $B_j\subseteq R_i$, and hence $U$ contains all right atoms $B_j$ such that $L_j$ is not a subset of $X$, and so $U$ contains $\bigcup_{\{j\colon L_j\not\subseteq X\}} B_j$.

On the other hand, if $L_j\subseteq X$ and $A_i\not\subseteq X$, then $A_i\not\subseteq L_j$, which by Proposition \ref{prop:r-quot}
gives that $B_j\not\subseteq R_i$.
Hence, $B_j\not\subseteq U$ if $L_j\subseteq X$, and so we get the reverse containment, establishing (1).

By Proposition \ref{prop:r-quot} we see that
for each left quotient $L_i$,
the inclusion $X\subseteq L_i$ holds if and only if
$B_i\subseteq \bigcap_{\{j\colon A_j\subseteq X\}}R_j$ holds and so we obtain (2).

The proofs of (3) and (4) are done similarly to (1) and (2).
$\square$
\vskip 2mm

A convenient tool for capturing much of this information 
comes from the \emph{quotient-atom} matrix \cite{KaWe70,Tamm16}.  If we adopt 
the notation of Notation \ref{notn:notn}, then this matrix is 
the $n\times m$ zero-one matrix whose $(i,j)$-entry (where we start our indices at zero) is $1$ exactly when $i\in S$, 
where $S\subseteq \{0,1,\ldots ,n-1\}$ is the set giving the 
left atom $A_j$ as a left atomic intersection 
$I_S$.  Equivalently, this is the case when $A_j\subseteq L_i$. 

In the case that $L$ is the regular language $\{\epsilon,a,a^2,ba\}$, the left quotients and left atoms are given in 
Equations (\ref{eq:leftquotient}) and (\ref{eq:leftatom}), and 
the expressions for left atoms as left atomic intersections are given in Equations (\ref{i1})--(\ref{i4}).  Using these data, we see that the quotient-atom matrix for $L=\{\epsilon,a,a^2,ba\}$ is given in Figure \ref{fig:qa}.
\begin{figure}[h]
   \begin{equation}
\left( \begin{array}{ccccc}
0 & 1 & 1 & 1 \\
0 & 0 & 1 & 1 \\
0 & 0 & 1 & 0 \\
0 & 0 & 0 & 1 \\
0 & 0 & 0 & 0
\end{array}\right)
\end{equation}
    \caption{The quotient-atom matrix for $L=\{\epsilon,a,a^2,ba\}$.}
    \label{fig:qa}
\end{figure}

 We note that one can do an analogous construction with right atoms and right quotients and one will then obtain the transpose of the quotient-atom matrix.  The quotient-atom matrix allows one to understand non-empty intersections of non-empty sets of left and right quotients in terms of maximal grids of the {quotient-atom matrix} \cite{KaWe70,Tamm16}.

\section{The isomorphism ${\rm Latt}(L,\cup, \mathsf{L})\cong {\rm Latt}(L,\cup, \mathsf{R})^*$}
\label{sec:proofa}
In this section, we give the proof of 
Theorem \ref{thm:main}(a). 

We define a set map
\begin{equation}
\Psi: {\rm Latt}(L,\cup, \mathsf{L})\to {\rm Latt}(L,\cup, \mathsf{R})
\end{equation} by declaring that for $X\in  {\rm Latt}(L,\cup, \mathsf{L})$,
\begin{equation}
\Psi(X):= \bigcup_{\{i\colon A_i\not\subseteq X\}} R_i.
\label{eq:Psi}
\end{equation}
We can similarly define a map 
\begin{equation}
\Psi': {\rm Latt}(L,\cup, \mathsf{R})\to {\rm Latt}(L,\cup, \mathsf{L}),
\end{equation}
where for $Y\in {\rm Latt}(L,\cup,\mathsf{R})$, we define
\begin{equation}
\Psi'(Y):= \bigcup_{\{i\colon B_i\not\subseteq Y\}} L_i.
\label{eq:Psi'}
\end{equation} 
We shall show that the maps $\Psi$ and $\Psi'$ are inverses of each other and that $\Psi$ induces a lattice isomorphism between 
${\rm Latt}(L,\cup, \mathsf{L})$ and ${\rm Latt}(L,\cup, \mathsf{R})^*$.

To continue with the example when $L=\{\epsilon,a,a^2,ba\}$, it can be checked that the map $\Psi$ is defined by the assignments
$\Psi(\varnothing)= \{\epsilon,a,a^2,b,ba\}$,
$\Psi(\{\epsilon\})=\{\epsilon,a,b\}$, $\Psi(\{a\})=\{\epsilon,a,a^2,ba\}$, $\Psi( \{\epsilon,a\})=\{\epsilon\}$, and $\Psi(\{\epsilon,a,a^2,ba\})= \varnothing$, which is capturing the dual structure of the lattices in Figure \ref{fig:L2}.

\begin{lemma} Let $\Psi$ and $\Psi'$ be the maps defined in Equations (\ref{eq:Psi}) and (\ref{eq:Psi'}).  Then $\Psi$ and $\Psi'$ are set-theoretic inverses of each other.
\label{lem:inverse}
\end{lemma}
\emph{Proof.} Let
$X\in \mathrm{Latt}(L,\cup, \mathsf{L})$.  Then $Y:=\Psi(X)$ is the union of all $R_i$ such that $A_i\not\subseteq X$.  Using Lemma \ref{prop:U-V} we then see
\begin{equation}\label{eq:PsiY}
\Psi(X) = \bigcup_{\{j\colon L_j\not\subseteq X\}} B_j.
\end{equation}
Then $\Psi'(Y)=\bigcup_{B_k\not\subseteq Y} L_k$.
Since right atoms are disjoint, from Equation (\ref{eq:PsiY}) we see that 
$B_j$ is not a subset of $Y$ if and only if $L_j\subseteq X$. Thus $\Psi'(Y)$ is the union of all left quotients $L_j$ contained in $X$, which is precisely $X$ as $X$ is a union of left quotients. 

The fact that $\Psi\circ \Psi'$ is the identity of ${\rm Latt}(L,\cup, \mathsf{R})$ is proved with a symmetric argument, again using Lemma \ref{prop:U-V}.
$\square$

\begin{lemma} Let $\Psi$ and $\Psi'$ be the maps defined in Equations (\ref{eq:Psi}) and (\ref{eq:Psi'}).  
If $U_1$ and $U_2$ are unions of left quotients of $L$, 
then the following hold:
\begin{enumerate}
\item[(1)] $U_1\subseteq U_2\iff \Psi(U_2)\subseteq \Psi(U_1)$;
\item[(2)] $\Psi(U_1\cup U_2)$ is the largest union of right quotients that is contained in the intersection $\Psi(U_1)\cap \Psi(U_2)$;
\item[(3)] if $V$ is the largest union of left quotients contained in $U_1\cap U_2$, then $\Psi(V) = \Psi(U_1)\cup \Psi(U_2)$.
\end{enumerate}
\label{prop:inclusion} 
\end{lemma}
\emph{Proof.} 
It is immediate from the definition that if $U_1\subseteq U_2$ then $\Psi(U_2)\subseteq \Psi(U_1)$.  Similarly,
if $V_1$ and $V_2$ are unions of right quotients of 
$L$, then if $V_2\subseteq V_1$ then
$\Psi'(V_1)\subseteq \Psi'(V_2)$.  Taking $V_i=\Psi(U_i)$ for 
$i=1,2$, by Lemma \ref{lem:inverse}, if $\Psi(U_2)\subseteq \Psi(U_1)$ 
then $U_1\subseteq U_2$, which establishes (1).

To see (2), observe that since $\Psi$ reverses inclusions, we have
$\Psi(U_1\cup U_2)\subseteq \Psi(U_1)\cap \Psi(U_2)$.
Now suppose that $R_i$ is a right quotient that is contained in $\Psi(U_1)\cap \Psi(U_2)$.  Then since $\Psi$ reverses inclusions and $\Psi'$ is the inverse of $\Psi$, we have that $\Psi'$ also reverses inclusions and so 
$\Psi'(R_i)\supseteq U_1$ since $R_i\subseteq \Psi(U_1)$, and similarly $\Psi'(R_i)\supseteq U_2$.
Hence $\Psi'(R_i)\supseteq U_1\cup U_2$ and so applying $\Psi$ and using once more that it reverses inclusions, we see that
$R_i$ is contained in $\Psi(U_1\cup U_2)$.  Thus $\Psi(U_1\cup U_2)$ is the largest union of right quotients contained in $\Psi(U_1)\cap \Psi(U_2)$, which shows (2).

We now prove (3).
Let $V$ be the union of all left quotients contained in $U_1\cap U_2$.  Then since $\Psi$ reverses inclusions, we have $\Psi(V)\supseteq \Psi(U_1)$ and similarly $\Psi(V)\supseteq \Psi(U_2)$, which shows that $\Psi(V)\supseteq \Psi(U_1)\cup \Psi(U_2)$.  To show equality, notice that if 
$\Psi(V)$ strictly contains $\Psi(U_1)\cup \Psi(U_2)$, then there is some right atom $B_k$ contained in $\Psi(V)$ that is neither contained in $\Psi(U_1)$ nor in $\Psi(U_2)$.  
Then since $B_k\subseteq \Psi(V)$, we have 
$L_k\not\subseteq V$ by Equation (\ref{eq:Psi}) and Lemma \ref{prop:U-V}.  But the fact that $B_k$ is not contained in $\Psi(U_1)$ gives that $L_k\subseteq U_1$ and similarly $L_k\subseteq U_2$.  Hence $L_k\subseteq U_1\cap U_2$.  But this contradicts the fact that we chose $V$ to be the union of left quotients contained in $U_1\cap U_2$. Thus we get (3).  
 $\square$

\vskip 2mm
\noindent \emph{Proof of Theorem \ref{thm:main}(a).} 
The fact that $\Psi$ gives a poset isomorphism between the lattice $
{\rm Latt}(L,\cup,\mathsf{L})$ and the dual lattice ${\rm Latt}(L,\cup,
\mathsf{R})^*$ follows from Lemmas \ref{lem:inverse} and 
\ref{prop:inclusion} (1).  Lemma \ref{prop:inclusion} (2) and 
(3) show that $\Psi$ preserves respectively the meet and join operations on these posets, as described in the definitions.  
$\square$
\section{The isomorphism ${\rm Latt}(L,\cap, \mathsf{L})\cong {\rm Latt}(L,\cap, \mathsf{R})^*$}
\label{sec:proofb}
The aim of this section is to prove 
Theorem \ref{thm:main}(b) involving intersections of left and right quotients.  

We now define maps
\begin{equation}
\Phi: {\rm Latt}(L,\cap, \mathsf{L})\to
{\rm Latt}(L,\cap, \mathsf{R})
\end{equation} and
\begin{equation}
\Phi' :{\rm Latt}(L,\cap, \mathsf{R})\to
{\rm Latt}(L,\cap, \mathsf{L})
\end{equation} as follows.
If $X\in {\rm Latt}(L,\cap,\mathsf{L})$, we define
\begin{equation}\Phi(X) = \bigcap_{A_j\subseteq X} R_j,
\label{eq:Phi}
\end{equation} and if $Y$ is an intersection of right quotients of $L$, we define
\begin{equation}\Phi'(Y) = \bigcap_{B_j\subseteq Y} L_j.
\label{eq:Phi'}
\end{equation}
The following lemmas can be proved in a similar manner to the method of proof for Lemmas \ref{lem:inverse} and \ref{prop:inclusion}.
\begin{lemma} Let $\Phi$ and $\Phi'$ be the maps defined in Equations (\ref{eq:Phi}) and (\ref{eq:Phi'}).  Then $\Phi$ and $\Phi'$ are set-theoretic inverses of each other.
\label{lem:inverse2}
\end{lemma}

\begin{lemma}
\label{prop:inclusion2}
Let $\Phi$ and $\Phi'$ be the maps defined in Equations (\ref{eq:Phi}) and (\ref{eq:Phi'}). 
If $U_1$ and $U_2$ are intersections of left quotients of $L$, then the following hold:
\begin{enumerate}
\item[(1)] $U_1\subseteq U_2\iff \Phi(U_2)\subseteq \Phi(U_1)$;
\item[(2)] $\Phi(U_1\cap U_2)$ is the smallest intersection of right quotients that contains the union $\Phi(U_1)\cup \Phi(U_2)$;
\item[(3)] if $V$ is the smallest intersection of left quotients that contains $U_1\cup U_2$, then $\Phi(V) = \Phi(U_1)\cap \Phi(U_2)$.
\end{enumerate}
\end{lemma}

\vskip 2mm
\noindent \emph{Proof of Theorem \ref{thm:main}(b).} 
This is proved similarly to Theorem \ref{thm:main}(a), but where we now use Lemmas \ref{lem:inverse2} and \ref{prop:inclusion2}.
$\square$

\section{Semimodules and semilattices}
\label{sec:semi}
In this section, we reinterpret our results algebraically and note connections with work of Im and Khovanov \cite{SeongIm}.

Let $\mathbb{B}$ denote the Boolean semiring, which is the set $\{0,1\}$ endowed with binary operations $+$ and $\cdot$ as in the tables from Figure \ref{fig:T}.

\begin{figure}[h]
    \centering
    \begin{subfigure}[t]{0.45\linewidth}
        \centering
        \begin{tabular}{c|cc}
            $+$ & 0 & 1 \\
            \hline
            0 & 0 & 1 \\
            1 & 1 & 1 \\
        \end{tabular}
        \label{fig:addition}
    \end{subfigure}
    \hfill
    \begin{subfigure}[t]{0.45\linewidth}
        \centering
        \begin{tabular}{c|cc}
            $\cdot$ & 0 & 1 \\
            \hline
            0 & 0 & 0 \\
            1 & 0 & 1 \\
        \end{tabular}
        \label{fig:multiplication}
    \end{subfigure}
    \caption{Addition and multiplication tables for the Boolean ring $\mathbb{B}$.}
    \label{fig:T}
\end{figure}

    A Boolean semimodule $M$ is a commutative monoid 
    (written additively and with an identity element $0_M$) equipped with 
    a scalar multiplication map $$\cdot: \mathbb{B}\times M 
    \to M$$ satisfying
    
    $1\cdot m=m$ for all $m\in M$, $0\cdot m=0_M$ for all $m\in M$ and $b\cdot (m+n)=b\cdot m+ b\cdot n$ and $(b+c)\cdot m=b\cdot m + c\cdot m$ for all $b,c\in \mathbb{B}$ and all $m,n\in M$.  
    
    In particular, if $M$ is a Boolean semimodule then for $m\in M$ we have $m+m = (1+1)\cdot m = 1\cdot m =m$, and so all elements of $M$ are idempotent.  A Boolean semimodule can be viewed as a join-semilattice (that is a partially ordered set in which any two elements have a least upper bound) as follows.  Given a Boolean semimodule $M$ we can define a partial order $\le$ on $M$ by declaring that $m\le n$ whenever $m+n=n$.  We can then define a join operation on $M$ by declaring that $m\vee n:=m+n$.  It is straightforward to check that this gives $M$ the structure of a join semilattice. Conversely, given a join semilattice $\Lambda$ with a least element $m_0$, one can endow $\Lambda$ with the structure of a Boolean semimodule by taking the join operation to be addition and taking $m_0$ to be the zero element.
    In case $M$ is a finite semimodule, $M$ is in fact a lattice with meet defined by taking
$m\wedge n$ to be the join of all elements $q$ that are less than or equal to both $m$ and $n$, and with unique maximal element given by taking the join of all elements of the lattice.

Given a Boolean semimodule $M$, one has a dual module 
$M^* = {\rm Hom}_{\mathbb{B}}(M,\mathbb{B})$, where ${\rm Hom}_{\mathbb{B}}(M,\mathbb{B})$ is the set of $\mathbb{B}$-linear maps from $M$ to $\mathbb{B}$.  We observe that $M^*$ is itself a Boolean semimodule, since we can add maps and have a zero map.   We then have a natural $\mathbb{B}$-bilinear pairing $\langle\, , \, \rangle: M\times M^*\to \mathbb{B}$ given by $\langle m,f\rangle = f(m)$ for $m\in M$ and $f\in M^*$.  
For a finite Boolean semimodule $M$, viewed as a semilattice, $M^*$ is just the dual semilattice of $M$. 

We note that for a finite alphabet $\Sigma$, we can construct the Boolean lattice ${\rm Bool}(\Sigma):=2^{\Sigma^*}$, consisting of subsets of $\Sigma^*$ partially ordered by inclusion and where meet and join are given by intersection and union respectively. 
Then given a regular language $L\subseteq \Sigma^*$, we have a $\mathbb{B}$-bilinear map, which we call the \emph{Im-Khovanov pairing} with respect to $L$,
$$\langle ~ ,~\rangle_L: {\rm Bool}(\Sigma)\times {\rm Bool}(\Sigma)\to \mathbb{B}$$
defined by 
\begin{equation}
\langle A,B\rangle_L = \left\{ \begin{array}{ll} 1 & \mathrm{if~there~exists}~w\in A, v\in B~{\rm such~that}~wv\in L;\\
0 & \mathrm{otherwise},\end{array}\right.
\end{equation}
for $A,B\subseteq \Sigma^*$.
From its definition, this is easily seen to be $\mathbb{B}$-bilinear and this pairing appears in the work of Im and Khovanov \cite[\S4]{SeongIm}.

For the remainder of this section, we adopt the notation of Notation \ref{notn:notn} and let $\langle ~ , ~ \rangle$ denote the Im-Khovanov pairing with respect to $L$. Then by Proposition \ref{prop:atoms-rquot},
$B_i$ is the left language of a state $q_i$ of the minimal DFA accepting $L$,
and $L_i$ is the corresponding right language of $q_i$.
Therefore, $\langle B_i,A_j \rangle =1$ if and only if $A_j \cap L_i \ne \varnothing$.
Similarly, using the átomaton of $L$, we obtain that the property $\langle B_i,A_j \rangle =1$ is equivalent to $B_i \cap R_j \ne \varnothing$.
On the other hand, left quotients are unions of left atoms and left atoms are disjoint, and so $A_j\cap L_i$ is non-empty if and only if $A_j\subseteq L_i$, and we have an analogous fact for right quotients and right atoms.
Hence, we have the equivalences
\begin{equation}
\label{eq:pairing1}
\langle B_i, A_j\rangle=1 \iff B_i\subseteq R_j \iff A_j\subseteq L_i,
\end{equation}
which can be thought of as an algebraic reformulation of Proposition \ref{prop:r-quot}. In general, if $X$ is a union of left atoms, then we have
$\langle B_i, X\rangle =0\iff X\cap L_i =\varnothing$, and if $Z$ is a union of right atoms, we have
$\langle Z, A_j\rangle =0 \iff Z\cap R_j=\varnothing$.

The pairing $\langle ~, ~ \rangle$ restricts to pairings 
$$\langle ~ ,~ \rangle: \mathrm{Latt}(L,\cup, \mathsf{R}) \times 
\mathrm{Latt}(L,\cup, \mathsf{L})\to \mathbb{B}$$
and
$$\langle \, ,\, \rangle: \mathrm{Latt}(L,\cap, \mathsf{R}) \times 
\mathrm{Latt}(L,\cap, \mathsf{L})\to \mathbb{B}.$$
We now give a description of the maps $\Psi$ and $\Phi$ from Equations (\ref{eq:Psi}) and (\ref{eq:Phi}) in terms of the Im-Khovanov pairing.  In order to express this, for a subset $Y$ of $\Sigma^*$, we let $Y^{\perp}$ denote the \emph{orthogonal complement} of $Y$, which is the subset of $\Sigma^*$ consisting of words $w$ with the property that $\langle Y,\{w\}\rangle=0$.
\begin{proposition} Let $\Psi$ and $\Phi$ be the maps given in Equations (\ref{eq:Psi}) and (\ref{eq:Phi}).  Then we have the following:
\begin{enumerate}
\item[(1)] for $X$ a union of left quotients,
$\Psi(X) = \bigcup_{\{k \colon \langle B_k, \ol{X}\rangle =1\}} B_k$;
\item[(2)]  for $Z$ an intersection of left quotients, 
$\Phi(Z) = \bigcup_{\{k~ \colon ~Z\cap B_k^{\perp}=\varnothing\}} B_k.$
\end{enumerate}
\end{proposition}
\emph{Proof.} Let $X$ be a union of left quotients.
Then by Equation (\ref{eq:Psi}), $$\Psi(X) = \bigcup_{\{j\colon A_j\not\subseteq X\}} R_j.$$  Since each right quotient is a union of right atoms, and since right atoms are disjoint, we see that $\Psi(X)$ is uniquely expressible as a union of right atoms.  Then $B_k\subseteq \Psi(X)$ if and only if there is some $j$ such that $A_j\not\subseteq X$ and $B_k\subseteq R_j$.  Notice that since $X$ is a union of left atoms, $A_j\not\subseteq X$ if and only if $A_j\subseteq \ol{X}$, and so we see by Equation (\ref{eq:pairing1}) that
$B_k\subseteq \Psi(X)$ if and only if there is an index $j$ such that $A_j\subseteq \ol{X}$ and $\langle B_k, A_j\rangle =1$.  Finally, bilinearity of our pairing says that 
$$B_k\subseteq \Psi(X) \iff \langle B_k, \ol{X}\rangle =1.$$
This completes the proof of (1).


Next let $Z$ be an intersection of left quotients.  
Then by Equation (\ref{eq:Phi}) we have 
$$\Phi(Z)= \bigcap_{\{j\colon A_j\subseteq Z\}} R_j.$$
 Notice that since right atoms are disjoint and since each right quotient is a union of right atoms, $B_k\subseteq \Phi(Z)$ if and only if 
$B_k\subseteq R_j$ for all $j$ such that $A_j\subseteq Z$.  Again, by 
Equation (\ref{eq:pairing1}), this is equivalent to 
$\langle B_k, A_j\rangle =1$ for all $j$ such that $A_j\subseteq Z$.  Notice that $\langle B_k, A_j\rangle =1$ if and only if $A_j$ is completely contained in $L_k$, and hence if $\langle B_k, A_j\rangle =1$, then $\langle B_k, Y\rangle =1$ for all non-empty subsets $Y$ of $A_j$.  Hence this is equivalent to saying that $Z$ does not intersect the orthogonal complement of $B_k$, and so the result follows.
$\square$
\vskip 2mm

One can also interpret the quotient-atom matrix in terms of the Im-Khovanov pairing, if one views the entries of the matrix as living in the Boolean semiring $\mathbb{B}$.  For the quotient-atom matrix, the $(i,j)$-entry is $1$ if $L_i$ appears in the atomic intersection giving $A_j$.  Equivalently, the $(i,j)$-entry is $1$ precisely when 
$A_j\subseteq L_i$, which by Equation (\ref{eq:pairing1}) occurs precisely when $\langle B_i, A_j\rangle =1$.  In particular, we have the following reinterpretation of the quotient-atom matrix.

\begin{proposition}
The quotient-atom matrix is the $n\times m$ matrix whose $(i,j)$-entry is $\langle B_i,A_j\rangle$.
\end{proposition}

\section{Complexity}
\label{sec:complexity}
In this section, we look at when the lattices we construct can be in some sense as large as possible. 

If we adopt the notation of Notation \ref{notn:notn}, then there are at most $2^n$ unions of left quotients and at most $2^m$ unions
of right quotients of $L$. By Theorem \ref{thm:main}, the number of unions of left quotients is equal to
the number of unions of right quotients, and hence there are at most $2^{\min(m,n)}$
unions of left/right quotients of $L$.

It is also not difficult to see that if $L$ has $2^n-1$ positive atoms---that is,
all possible positive atoms exist---then there are $2^n$ unions of
left quotients.
We show, however, to realize this maximal complexity, only $n$ left atoms of $L$ are required.

\begin{proposition}
\label{prop:union-complexity}
There are $2^n$ unions of left quotients of $L$ if and only if
all the left atomic intersections with one uncomplemented and
$n-1$ complemented left quotients are non-empty.
\end{proposition}
\emph{Proof.}
Let us suppose that all the left atomic intersections with one uncomplemented and
$n-1$ complemented left quotients of $L$ are non-empty.
That is, for every $i\in\{0,\ldots,n-1\}$, the left atomic intersection $I_{\{i\}}$ is non-empty.
Hence, for each left quotient $L_{i}$, there is at least one atom,
namely $I_{\{i\}}$, contained in $L_{i}$ and not contained in any other left quotient.  Since the left atoms are pairwise disjoint, this implies that there are $2^n$ distinct unions of left quotients of $L$.

Conversely, if $I_{\{i\}}$ is empty for some $i$, then it is easily checked that
$$\bigcup_{j\neq i} L_j = \bigcup_{j} L_j$$
and so the number of unions of left quotients of $L$ is strictly less than $2^n$.
$\square$
\vskip 2mm

A similar result can be achieved for the complexity of intersections of
left quotients of $L$. 
\begin{proposition}
\label{prop:intersec-complexity}
There are $2^n$ intersections of left quotients of $L$ if and only if
all the left atomic intersections with $n-1$ uncomplemented and
one complemented left quotients are non-empty.
\end{proposition}
\emph{Proof.}
First, let us assume that for every $k\in\{0,\ldots,n-1\}$, $Z_k:=I_{\{0,\ldots ,n-1\}\setminus \{k\}}$ is non-empty.

Now, consider any intersection of left quotients
$X=L_{i_1}\cap \cdots \cap L_{i_s}$.  Then one can verify that $Z_k\subseteq X$ if and only if $k\not\in \{i_1,\ldots ,i_s\}$ for $k\in \{0,\ldots ,n-1\}$.
Thus by checking which of the left atoms $Z_0,\ldots ,Z_{n-1}$ are subsets of an intersection of left quotients, we can uniquely recover the left quotients appearing in the intersection and so we obtain $2^n$ distinct intersections. 

Conversely, suppose that 
for some $k$, the intersection $Z_k$ is empty.  Then $\bigcap_{j\neq k}L_j$ has empty intersection with $\ol{L_k}$ and thus is contained in $L_k$.  Hence
$$\bigcap_{j\neq k} L_j = \bigcap_{j} L_j,$$ and so the number of intersections of left quotients of $L$ is strictly smaller than $2^n$. $\square$
\vskip 2mm
We get the following result as an immediate consequence of Propositions \ref{prop:union-complexity} and \ref{prop:intersec-complexity}.
\begin{corollary}
\label{cor:cor}
Let $n>2$. Then the $2n$ atoms of $L$, described in
Propositions~\ref{prop:union-complexity} and \ref{prop:intersec-complexity},
are necessary and sufficient to obtain the equalities
$$\left| {\rm Latt}(L,\cup, \mathsf{L})\right| =
\left| {\rm Latt}(L,\cap, \mathsf{L})\right| = 2^n.$$
\end{corollary}

\section{Conclusions and further work}
\label{sec:concl}

Corollary \ref{cor:cor} gives an efficient means for checking that the lattices we obtain are of maximal possible complexity.  It would be interesting to know whether other lattice-theoretic properties for the lattices we consider can be efficiently checked or even characterized in terms of the associated automata.  
Of particular interest is the question of when our lattices are distributive.  In the framework considered by Im and Khovanov \cite{SeongIm}, the distributive property is key for associating topological quantum field theories to regular languages.

\bibliographystyle{eptcs}

\end{document}